%% file: main.tex
\documentclass[epj]{svjour}
\usepackage{fontenc}
\usepackage[utf8]{inputenc}
\usepackage[english]{babel}
\usepackage[fleqn]{amsmath}
\usepackage{mathtools,amssymb,bm}
\usepackage[retain-unity-mantissa = false]{siunitx}
\usepackage{slashed}
\usepackage{graphicx}
\usepackage{placeins}
\usepackage{tabu}
\usepackage{multirow}
\usepackage{xcolor}
\usepackage[backend=bibtex,style=phys,biblabel=brackets,pageranges=false,
url=true,maxbibnames=6]{biblatex}

\renewbibmacro{in:}{}
\DeclareFieldFormat[article]{journaltitle}{\textit{#1}}
\AtEveryBibitem{\clearlist{language}}
\AtEveryBibitem{\clearfield{title}}
\DeclareBibliographyCategory{needsurl}
\renewbibmacro*{url+urldate}{
	\ifcategory{needsurl}{
		\printfield{url}
		\iffieldundef{urlyear}{}{
			\setunit*{\addspace}
			\printurldate
		}
	}{}
}
\addtocategory{needsurl}{oppitz_master_2017}

\input{pgfsetup.tex}

\newcommand{\diff}{\mathrm d}
\newcommand*{\eh}[1]{\mathrm e^{#1}}
\newcommand{\eperp}{\epsilon_\perp}
\newcommand{\tramp}{t_\text{ramp}}
\newcommand{\tderamp}{t_\text{deramp}}
\newcommand{\tflat}{t_\text{flat}}
\newcommand{\tr}{t_\text{rel}}
\newcommand{\tm}{t_\text{max}}

\begin{document}

\title{Assisted Vacuum Decay by Time Dependent Electric Fields}
\author{A.~Otto\inst{1,2} \and H.~Oppitz\inst{2} \and B.~K\"ampfer\inst{1,2}}

\institute{Institute of Radiation Physics, Helmholtz-Zentrum Dresden-Rossendorf,
01328 Dresden, Germany \and Institut f\"ur Theoretische Physik, Technische
Universit\"at Dresden, 01068 Dresden, Germany}

\date{Received: date / Revised version: date}

\abstract{
We consider the vacuum decay by electron-positron pair production in spatially
homogeneous, time dependent electric fields by means of quantum kinetic
equations.
Our focus is on the impact of various pulse shapes as envelopes of oscillating
fields and the assistance effects in multi-scale fields, which are also seen in
photons accompanying the creation and motion of pairs.
}

\maketitle

\onecolumn
\section{Introduction}
Besides pair (e.g.\ electron-positron, $e^+$$e^-$) creation in counter
propagating null fields, e.g.\ by the Breit-Wheeler process, also other purely
electromagnetic fields qualify for pair production.
An outstanding example is the Schwinger effect originally meaning the
instability of a spatially homogeneous, purely electric field with respect to
the decay into a state with pairs and a screened electric field~\cite{schwinger}
(cf.~\cite{gelis_schwinger_2016} for a recent review).
The pair creation rate $\propto \exp\{ - \pi E_c / \vert \vec E \vert \}$ for
electric fields $\vec E$ attainable presently in mesoscopic laboratory
installations is exceedingly small since the Sauter-Schwinger (critical) field
strength~\cite{sauter} $E_c = m^2/|e| = \SI{1.3e16}{V/cm}$ for
electrons/positrons with masses $m$ and charges $\pm e$ is so large (we employ
here natural units with $c = \hbar = 1$).

An analogous situation of vacuum decay is met in the case, where a spatially
homogeneous electric field has a time dependence.
The particular case of a periodic field is dealt with in~\cite{brezin_pair_1970}
with the motivation that tightly focused laser beams can provide high field
strengths.
The superposition of a few laser beams, as considered, e.g.\ in~\cite
{narozhny_pair_2004}, can enlarge the pair yield noticeably.
A particular variant is the superposition of strong optical laser beams and
weaker but high-frequency (XFEL) beams.
If the frequency of the first field is negligibly small while that of the second
field is sufficiently large, the tunneling path through the positron-electron
mass gap is shortened by the assistance of the multi-photon effect and, as a
consequence, the pair production is enhanced.
This is the dynamically assisted Schwinger process~\cite
{schutzhold_dynamically_2008,dunne_catalysis_2009}.
As (dynamically) assisted dynamical Schwinger effect one can denote the pair
creation (vacuum decay) where the time dependence of both fields matters,
cf.~\cite{akal_euclidean_2017}.
Many investigations in this context are constrained to spatially homogeneous
field models, that is to the homogeneity region of anti-nodes of pairwise
counter propagating and suitably polarized laser beams.
Accounting for spatial gradients is interesting, e.g.\ w.r.t.\ critical
suppression~\cite{gies_critical_2016,gies_critical_2017}.

Since the Coulomb fields accompanying heavy and super-heavy atomic nuclei or
ions in a near-by passage can achieve ${\cal O} (E_c)$, the vacuum break down
for such configurations with inhomogeneous static or slowly varying fields
has been studied extensively~\cite{greiner_3,rafelski_superheavy_1971,
muller_solution_1972,muller_solution_1973,fillion-gourdeau_enhanced_2013}, thus
triggering a wealth of follow-up investigations.
Other field combinations, e.g.\ the nuclear Coulomb field and XFEL/laser beams,
are also conceivable~\cite{augustin_nonlinear_2014,di_piazza_effect_2010}, but
will not be addressed here (cf.~\cite{di_piazza_extremely_2012} for a survey).

In the present note, we consider the pair production as vacuum decay in
spatially uniform, time dependent, external model fields.
As novelties we provide (i) the account of some generic pulse shapes and their
impact on the pair spectrum (section~2) and (ii) examples of assistance effects
which occur in multi-scale fields (sections~3 and~4).
The summary can be found in section 5.

\section{Schwinger pair production for various pulse shapes}
Even for the simple field model $E(t,\vec x) = E(t)$, i.e.\ a spatially
homogeneous but time dependent field, the problem remains highly non-linear with
intricate parameter dependence.
The quantum kinetic equations for the pair distribution $f(t,\vec p)$ as a
function of time $t$ and for a pair member with momentum $\vec p=(\vec p_\perp,
p_\parallel)$ can be cast in the form~\cite{schmidt_quantum_1998,
schmidt_non-markovian_1999}
\begin{align}
\begin{aligned}
\frac{\diff}{\diff t}\begin{pmatrix} f(t,\vec p)\\ u(t,\vec p)\\ v(t,\vec p)
\end{pmatrix} = \begin{pmatrix} 0 & Q(t,\vec p) & 0\\
-Q(t,\vec p) & 0 & -2\Omega(t,\vec p)\\
0 & 2\Omega(t,\vec p) & 0\end{pmatrix}
\begin{pmatrix} f(t,\vec p)\\ u(t,\vec p)\\ v(t,\vec p)\end{pmatrix} +
\begin{pmatrix} 0\\ Q(t,\vec p)\\ 0\end{pmatrix}
\end{aligned}
\label{QKE}
\end{align}
with
\begin{align}
Q(t,\vec p) = \frac{eE(t)\eperp}{\Omega(t,\vec p)^2},\quad
\eperp = \sqrt{m^2+p_\perp^2},\quad
\Omega(t,\vec p) = \sqrt{\eperp^2+\bigl(p_\parallel-eA(t)\bigr)^2}
\end{align}
and initial conditions $f(t_0) = u(t_0) = v(t_0) = 0$ and $t_0\to-\infty$.
The quantities $u$ and $v$ are fiducial ones, but might be considered as
describing certain aspects of vacuum polarization effects.
Our approach is based on the Dirac equation but works equally well for the
Klein-Gordon equation, i.e.\ bosons.

To arrive at some systematics, let us consider the type of field models
described by the four-potential
\begin{align}
A_t=A_x=A_y=0,\quad A(t)\equiv A_z(t) = K(t)h(\nu t)
\label{A}
\end{align}
yielding the electric field with components $E_x=E_y=0$, $E_z=-\partial A_z/
\partial t$ and vanishing magnetic field.
The periodic function $h(x)$, $x\equiv\nu t$, is for a multi-scale field with
basic frequency $\nu$
\begin{align}
h(x) = \sum_{i=1}^N \frac{E_i}{N_i\nu}\cos(N_ix+\varphi_i);
\label{h}
\end{align}
we put $N_1=1$ and discard the phase shifts $\varphi_i$, i.e.\ $\varphi_i=0$.
The dimensionless quantity $K(t)\in C^\infty$ is the pulse shape function
(formerly dubbed ``pulse envelope'', cf.~\cite{otto_lifting_2015}).
Its effect on pair creation was already investigated, e.g.\ in~\cite
{kohlfurst_optimizing_2013,linder_pulse_2015,aleksandrov_pulse_2017}, see
also~\cite{torgrimsson_dynamically_2017}.
We specify the pulse shape function now as
\begin{align}
K(t) = \begin{cases}
\eh{-(t/t_G)^{2n}}, & \text{Gauss, super-Gauss},\\
K_\text{f.t.}(t;\tflat,\tramp,\tderamp), & \text{flat-top pulse}.
\end{cases}
\label{pulse shapes}
\end{align}
That is, $n=1$ refers to a Gauss pulse shape, $n>1$ is a super-Gauss, and
$K_\text{f.t.}$ is exactly as in~\cite{otto_lifting_2015} with a flat-top time
interval $\tflat$ ($K(t\in\tflat)=1$), a ramping time $\tramp$ and deramping
time $\tderamp$ ($0\le K(t\in\tramp,\tderamp)\le1$) as well as prior-pulse and
posterior-pulse values of $K=0$.
The super-Gauss pulse has for sufficiently large values of $n$, e.g.\ $n>4$, a
near-flat top interval and far-extended (de-)ramping periods, as the Gauss
pulse.
We consider (i) not too short pulses, i.e.\ within the pulse the field
oscillates at least a few times (so that in fact the impact of $\varphi_i$
becomes subleading) and (ii) time-symmetric pulses, i.e.\ $\tderamp=\tramp$.
The quantities $t_G$ and $\{\tflat, \tramp, \tderamp\}$ introduce further time
scales, in addition to $(N_i \nu)^{-1}$.

Generalizing the double Fourier expansion method in~\cite{otto_dynamical_2015},
valid in the low-density approximation, $f(t,\vec p)\ll1$, which is equivalent
to a linearized Riccati equation (cf.~\cite{dumlu_interference_2011}), some
approximate formula can be derived~\cite{oppitz_master_2017}, where
$\tflat>\tramp$ is supposed:
\begin{align}
\mathcal W(p_\perp,p_\parallel=0,\tr,n_\ell) = \frac49\nu^2
\mathcal E(p_\perp) \mathcal K(p_\perp,n_\ell)
\tr^{2\left(1-\frac{1}{n_\ell+1}\right)},\quad
\tr=\begin{cases}t_G, & \text{Gauss, super-Gauss,}\\
\tflat, & \text{flat-top,} \end{cases}
\label{W}
\end{align}
where the quantity $\mathcal W$ describes the envelope of the residual spectrum
$\lim_{t\to\infty}f(t,\vec p)$ at $p_\parallel\equiv p_z=0$ as a function of
$p_\perp=\sqrt{p_x^2+p_y^2}$.
The number $n_\ell$ is the order of the first non-vanishing derivative of
$K$ in~\eqref{pulse shapes} at the maximum $\tm$ where $K=1$.
Explicitly,
\begin{align}
n_\ell = \begin{cases}
2n, & \text{Gauss, super-Gauss,}\\
\to\infty, & \text{flat-top pulse.}
\end{cases}
\end{align}
The function
\begin{align}
\mathcal E(p_\perp) = \exp\left\{-4\frac m\nu G(p_\perp)\right\},
\quad\text{with}\quad
G(p_\perp) = \int\limits_0^{x_0}\!\diff x\left(1+\frac{p_\perp}{m}+
\left[\sum_{i=1}^N\frac{\sinh N_ix}{\gamma_i}\right]\right)^\frac12,
\label{envelope}
\end{align}
Keldysh parameters $\gamma_i=\frac{E_c}{E_i}\frac{N_i\nu}{m}$ and $x_0$ as the
zero of $\mathcal F = 1+\frac{p_\perp}{m} + \left[\sum_{i=1}^N
\frac{\sinh N_ix}{\gamma_i}\right]$, i.e.\ $\mathcal F(x_0)=0$, is independent
of the pulse shape.
The pulse shape dependence enters in~\eqref{W} via
\begin{align}
\mathcal K(p_\perp,n_\ell) = 2\left(\frac{(n_\ell+1)!}{2\left|K^{(n_\ell)}(\tm)
e\frac{\diff}{\diff e}\Omega_0(p_\perp)\right|}\right)^\frac{2}{n_\ell+1}
\Gamma^2\!\left(\frac{n_\ell+2}{n_\ell+1}\right)
\cos^2\left(\frac{\pi}{2n_\ell+2}\right)
\label{K}
\end{align}
with
\begin{align}
\Omega_0(p_\perp) = \frac{1}{2\pi}\int\limits_0^{2\pi}\!\diff x
\sqrt{m^2+p_\perp^2+\bigl(eh(x)\bigr)^2}
\end{align}
the effective energy as Fourier zero-mode from~\cite{otto_dynamical_2015}, for
further details cf.~\cite{oppitz_master_2017}.
For the flat-top pulse, the quantity $\left[(n_\ell+1)!/|K^{(n_\ell)}(\tm)|
\right]^\frac{2}{n_\ell+1}$ in Eq.~\eqref{K} has to be understood in the limit
$n_\ell\to\infty$, as all the pulse's derivatives at the maximum are zero.
This yields eventually
\begin{align}
\mathcal W(p_\perp,p_\parallel=0,\tr) = \frac29\nu^2 \mathcal E(p_\perp)\times
\begin{cases}
3\left(\frac{3}{2\left|e\frac{\diff}{\diff e}\Omega_0(p_\perp)\right|}
\right)^{2/3} \Gamma^2\!\left(\frac{4}{3}\right)\mathcal J, &
\text{Gauss ($n=1$),}\\
4\left(\frac{11}{2\left|e\frac{\diff}{\diff e}\Omega_0(p_\perp)\right|}
\right)^{2/11} \Gamma^2\!\left(\frac{12}{11}\right)
\cos^2\left(\frac{\pi}{22}\right)\mathcal J, &
\text{super-Gauss ($n=5$),}\\
\mathcal J, & \text{flat-top pulse ($n\to\infty$).}\\
\end{cases}
\tag{\ref{W}'}
\label{W'}
\end{align}
The advantage of~\eqref{W} and~\eqref{W'} is that they are valid for quite
different pulses and allow for the estimate of the envelope of the spectrum on
the line $p_\parallel=0$.
We emphasize the different scalings of the spectra with the relevant envelope
time scales:
\begin{figure}
\centering
\begin{tikzpicture}
\begin{groupplot}[
group style={
	group size=3 by 1,
	xlabels at=edge bottom,
	xticklabels at=edge bottom,
	ylabels at=edge left,
	yticklabels at=edge left,
	horizontal sep=0pt
},
width=0.39\textwidth,
xlabel shift=-3,
ylabel shift=-5,
xmin=0,
xmax=1.5,
xtick={0, 0.3, 0.6, 0.9, 1.2, 1.5},
x tick label style={/pgf/number format/.cd, fixed, fixed zerofill, precision=1,
/tikz/.cd},
ymin=1e-12,
ymax=1e0,
ytick={1e-12, 1e-10, 1e-8, 1e-6, 1e-4, 1e-2, 1e0},
xlabel=$p_\perp/m$,
ylabel={$f(t_\infty,p_\perp,p_\parallel=0)$},
ymode=log,
yminorticks=false,
title style={yshift=-8},
]

\nextgroupplot[xtick={0, 0.3, 0.6, 0.9, 1.2},
title={$a_{e^-e^+}=\num{4e-5}\lambda_C^{-2}$}]
\addplot[color=black,line width=1] table
{data/cut_one_field/gauss_n=1_E1=0.2_tau=50.dat};
\addplot[color=red,dashed,line width=1] table
{data/cut_one_field/gauss_n=1_E1=0.2_tau=50_envelope.dat};

\nextgroupplot[xtick={0, 0.3, 0.6, 0.9, 1.2},
title={$a_{e^-e^+}=\num{1e-4}\lambda_C^{-2}$}]
\addplot[color=black,line width=1] table
{data/cut_one_field/supergauss_n=5_E1=0.2_tau=50.dat};
\addplot[color=red,dashed,line width=1] table
{data/cut_one_field/supergauss_n=5_E1=0.2_tau=50_envelope.dat};

\nextgroupplot[title={$a_{e^-e^+}=\num{2e-4}\lambda_C^{-2}$}]
\addplot[color=black,line width=1] table
{data/cut_one_field/flattop_E1=0.2_tau=50.dat};
\addplot[color=red,dashed,line width=1] table
{data/cut_one_field/flattop_E1=0.2_tau=50_envelope.dat};
\end{groupplot}
\end{tikzpicture}
\caption{Residual phase space distribution $f(p_\parallel=0)$ as a function of
transverse momentum $p_\perp$ for various pulse shapes (left panel: Gauss;
middle panel: super-Gauss; right panel: flat-top).
Parameters are $E_1=0.2E_c$, $\nu=0.02m$ and $n=1$, $\nu t_G=25\cdot2\pi$ (left
panel), $n=5$, $\nu t_G=25\cdot2\pi$ (middle panel), $\nu\tramp=5\cdot2\pi$,
$\nu\tflat=50\cdot2\pi$ (right panel).
As estimator of the density we use here $a_{e^-e^+}=2\pi\int\!\diff p_\perp
p_\perp f(p_\perp,p_\parallel=0)$ depicted on top of the panels;
the quantity $\lambda_C=\frac{h}{mc}$ is the Compton wavelength.}
\label{gauss super-gauss otto}
\end{figure}
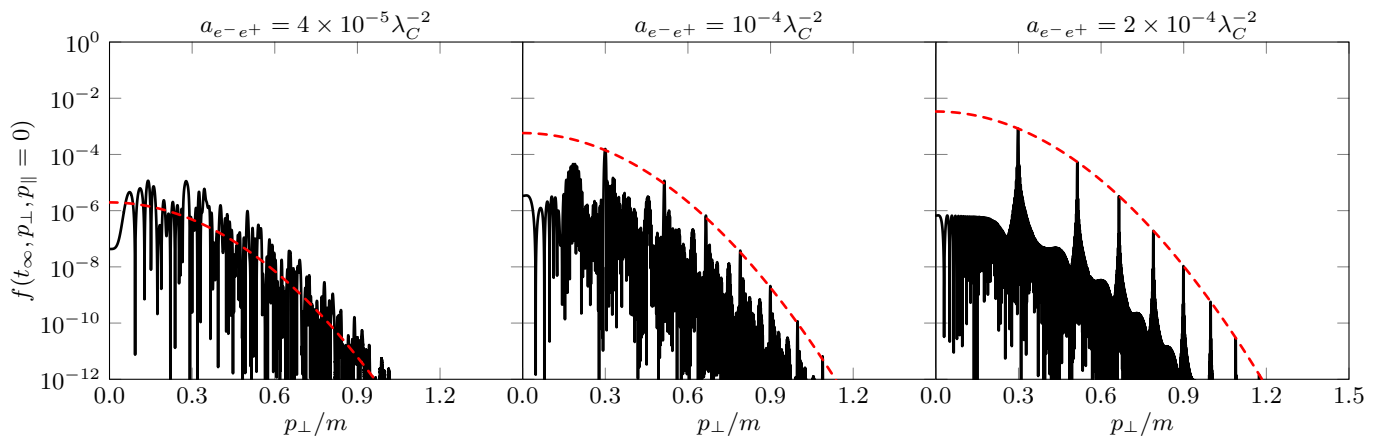
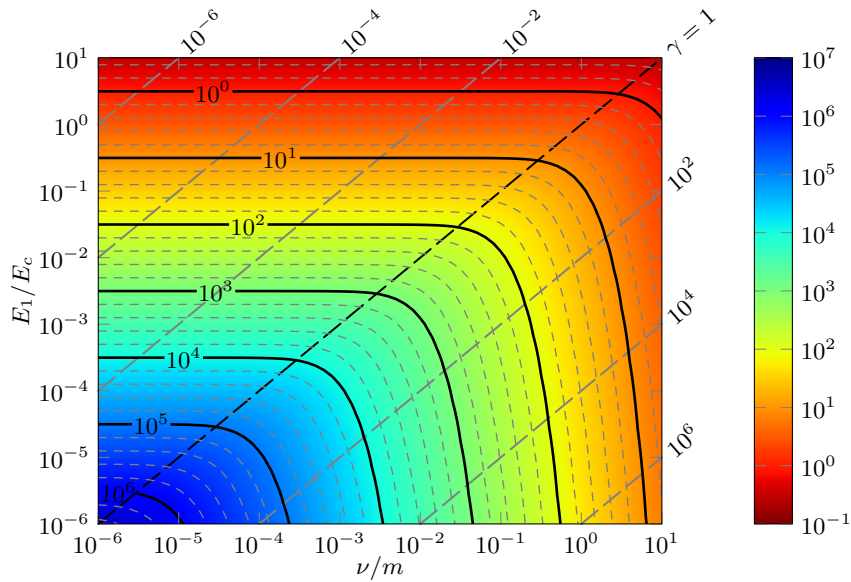
\begin{figure}
\newcommand{\datapath}{generated_data/env_exp_contour_lines/}
\centering
\begin{tikzpicture}
\begin{axis}[
width=0.5\textwidth,
xlabel shift=-5,
ylabel shift=-5,
xmin=1e-6,
xmax=1e1,
ymin=1e-6,
ymax=1e1,
xmode=log,
ymode=log,
xminorticks=false,
yminorticks=false,
xlabel=$\nu/m$,
ylabel=$E_1/E_c$,
point meta min=1e-1, point meta max=1e7,
colormap name=myjet_r,
colorbar,
colorbar style={ymode=log,yminorticks=false},
axis on top,
clip mode=individual,
]

\newcommand{\labelpos}{0.2}
\newcommand{\contourline}[1]{
\begin{scope}
\addplot[opacity=0,line width=1] table{\datapath 1e#1.0.dat}
node[pos=\labelpos,sloped,opacity=0,inner sep=0pt] (A) {$10^{#1}$};
\clip (A.north east) -- (A.north west) -- (A.south west) -- (A.south east) --
cycle (axis cs:1e-6,1e-6) rectangle (axis cs:10,10);
\addplot[opacity=1,line width=1] table{\datapath 1e#1.0.dat};
\end{scope}
\addplot[opacity=0,line width=1] table{\datapath 1e#1.0.dat}
node[pos=\labelpos,sloped,opacity=1,inner sep=0pt] {$10^{#1}$};
}

\addplot graphics[xmin=1e-6,xmax=1e1,ymin=1e-6,ymax=1e1] {generated_images/{env_exp_contour_range=1e-06_1e+01_1e-06_1e+01_1e-01_1e+07}.png};

\draw[line width=0.75,dash pattern=on 8pt off 3pt] (axis cs:1e-6,1e-6) --
(10, 10) node[right,rotate=40.7777777] {$\gamma=1$};

\draw[color=gray,line width=0.75,dash pattern=on 8pt off 3pt]
(axis cs:1e-4,1e-6) -- (10, 1e-1)
node[color=black,right,rotate=40.7777777] {$10^2$};
\draw[color=gray,line width=0.75,dash pattern=on 8pt off 3pt]
(axis cs:1e-2,1e-6) -- (10, 1e-3)
node[color=black,right,rotate=40.7777777] {$10^4$};
\draw[color=gray,line width=0.75,dash pattern=on 8pt off 3pt]
(axis cs:1,1e-6) -- (10, 1e-5)
node[color=black,right,rotate=40.7777777] {$10^6$};

\draw[color=gray,line width=0.75,dash pattern=on 8pt off 3pt]
(axis cs:1e-6,1e-4) -- (1e-1, 10)
node[color=black,right,rotate=40.7777777,yshift=2pt] {$10^{-2}$};
\draw[color=gray,line width=0.75,dash pattern=on 8pt off 3pt]
(axis cs:1e-6,1e-2) -- (1e-3, 10)
node[color=black,right,rotate=40.7777777,yshift=2pt] {$10^{-4}$};
\draw[color=gray,line width=0.75,dash pattern=on 8pt off 3pt]
(axis cs:1e-6,1) -- (1e-5, 10)
node[color=black,right,rotate=40.7777777,yshift=2pt] {$10^{-6}$};

\addplot[color=gray,line width=0.5,dashed] table{\datapath 1e-0.2.dat};
\addplot[color=gray,line width=0.5,dashed] table{\datapath 1e-0.4.dat};
\pgfplotsinvokeforeach{0,1,...,5}{
\contourline{#1}
\addplot[color=gray,line width=0.5,dashed] table{\datapath 1e#1.2.dat};
\addplot[color=gray,line width=0.5,dashed] table{\datapath 1e#1.4.dat};
\addplot[color=gray,line width=0.5,dashed] table{\datapath 1e#1.6.dat};
\addplot[color=gray,line width=0.5,dashed] table{\datapath 1e#1.8.dat};
}
\contourline{6}
\addplot[color=gray,line width=0.5,dashed] table{\datapath 1e6.2.dat};
\addplot[color=gray,line width=0.5,dashed] table{\datapath 1e6.4.dat};

\end{axis}
\end{tikzpicture}
\caption{Contour plot of $4\frac m\nu G$, i.e.\ the negative of the exponent of
$\mathcal E(\vec p=0)$ from Eq.~\eqref{envelope}, over the $E_1/E_c$
vs.\ $\nu/m$ plane.
The solid black curves are contour lines at integer powers of $10$, the grey
short dashed curves at $10^{0.2},\cdots,10^{0.8}$ between them.
The long dashed lines are the loci of constant Keldysh parameter
$\gamma=\gamma_1$.}
\label{landscape}
\end{figure}
\begin{align}
\mathcal J = \begin{cases}
t_G^{4/3}, & \text{Gauss ($n=1$),}\\
t_G^{20/11}, & \text{super-Gauss ($n=5$),}\\
\tflat^2, & \text{flat-top pulse ($n\to\infty$).}
\end{cases}
\end{align}

Figure~\ref{gauss super-gauss otto} exhibits three examples for the transverse
momentum spectrum at $p_\parallel=0$ for $N=1$, $E_1=0.2E_c$, $\nu=0.02m$ for a
Gauss pulse (left panel, $n=1$, $\nu t_G=25\cdot2\pi$) and a super-Gauss (middle
panel, $n=5$, $\nu t_G=25\cdot2\pi$) and a flat-top pulse (right panel,
$\nu\tflat=50\cdot2\pi$, $\nu\tramp=5\cdot2\pi$).
In fact,~\eqref{W'} delivers quite reliable estimates of the envelopes of the
spectra.
The latter ones require solutions of the quantum kinetic equations~\eqref{QKE}
for each value of $\vec p$ separately and with harsh numerical accuracy
requirements.
Thus,~\eqref{W} and~\eqref{W'} qualify for useful estimates of the maxima of the
spectral distributions for several relevant pulse shapes, e.g.\ by considering
$\mathcal W(p_\perp=0)$ as an estimator of the maximum peak height, supposed the
pulse is flat enough.
Obviously, the Gauss pulse is not flat enough; as a consequence, \eqref{W}
and~\eqref{W'} meet only the order of magnitude, at best.

To get a feeling for the parameter dependence of that maximum peak height, one
can inspect the pulse-shape independent quantity $\mathcal E(p_\perp=0)$.
Figure~\ref{landscape} exhibits the function $4\frac m\nu G(p_\perp=0,N=1)$ as
a contour plot over the $E_1/E_c$ vs.\ $\nu/m$ plane.
Since $\mathcal E=\exp\{-4\frac m\nu G\}$, large values of $4\frac m\nu G$
signal a strong suppression of the pair production.
The region left/above the Keldysh line $\gamma_1=1$ (heavy dashed line) is the
tunneling regime, where our above examples are located, while the region
right/below the Keldysh line $\gamma_1=1$ belongs to the multi-photon regime.

\section{Assisted Schwinger pair production}
\begin{figure}
\centering
\begin{tikzpicture}
\begin{groupplot}[
group style={
	group size=3 by 1,
	xlabels at=edge bottom,
	xticklabels at=edge bottom,
	ylabels at=edge left,
	yticklabels at=edge left,
	horizontal sep=0pt
},
width=0.39\textwidth,
xlabel shift=-3,
ylabel shift=-5,
xmin=0,
xmax=1.5,
xtick={0, 0.3, 0.6, 0.9, 1.2, 1.5},
x tick label style={/pgf/number format/.cd, fixed, fixed zerofill, precision=1,
/tikz/.cd},
ymin=1e-12,
ymax=1e0,
ytick={1e-12, 1e-10, 1e-8, 1e-6, 1e-4, 1e-2, 1e0},
xlabel=$p_\perp/m$,
ylabel={$f(t_\infty,p_\perp,p_\parallel=0)$},
ymode=log,
yminorticks=false,
title style={yshift=-8},
]

\nextgroupplot[xtick={0, 0.3, 0.6, 0.9, 1.2},
title={$a_{e^-e^+}=\num{6e-3}\lambda_C^{-2}$}]
\addplot[color=black,line width=1] table
{data/cut_two_fields/gauss_n=1_E1=0.2_tau=50_E2=0.05_N2=25.dat};
\addplot[color=red,dashed,line width=1] table
{data/cut_two_fields/gauss_n=1_E1=0.2_tau=50_E2=0.05_N2=25_envelope.dat};

\nextgroupplot[xtick={0, 0.3, 0.6, 0.9, 1.2},
title={$a_{e^-e^+}=\num{2e-2}\lambda_C^{-2}$}]
\addplot[color=black,line width=1] table
{data/cut_two_fields/supergauss_n=5_E1=0.2_tau=50_E2=0.05_N2=25.dat};
\addplot[color=red,dashed,line width=1] table
{data/cut_two_fields/supergauss_n=5_E1=0.2_tau=50_E2=0.05_N2=25_envelope.dat};

\nextgroupplot[title={$a_{e^-e^+}=\num{2e-2}\lambda_C^{-2}$}]
\addplot[color=black,line width=1] table
{data/cut_two_fields/flattop_E1=0.2_tau=50_E2=0.05_N2=25.dat};
\addplot[color=red,dashed,line width=1] table
{data/cut_two_fields/flattop_E1=0.2_tau=50_E2=0.05_N2=25_envelope.dat};
\end{groupplot}
\end{tikzpicture}
\caption{As in Fig.~\ref{gauss super-gauss otto} but for the superposition of
a second field with parameters $E_2=0.05E_c$, $N_2=25$.}
\label{gauss super-gauss otto ass}
\end{figure}
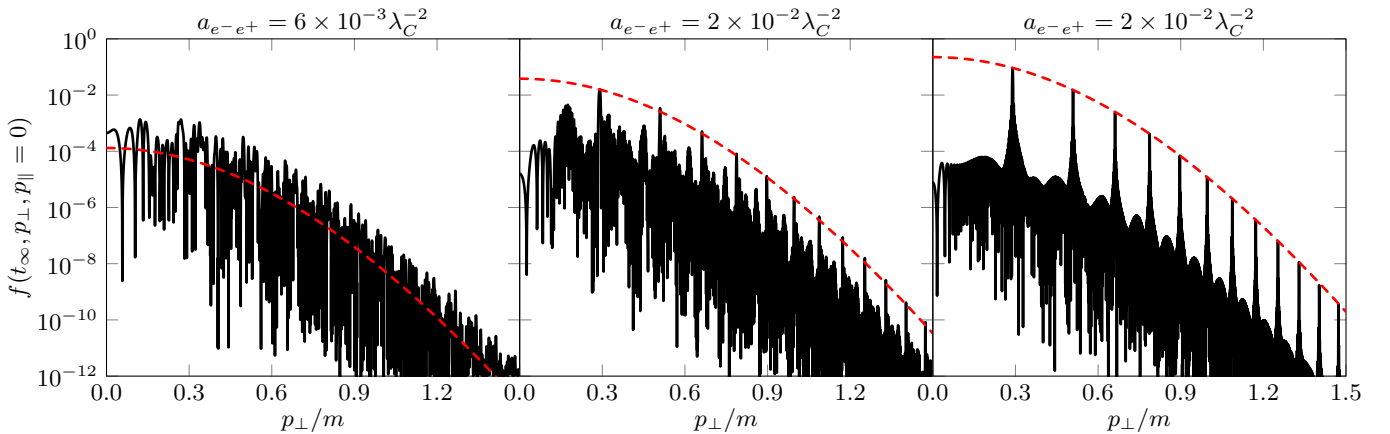
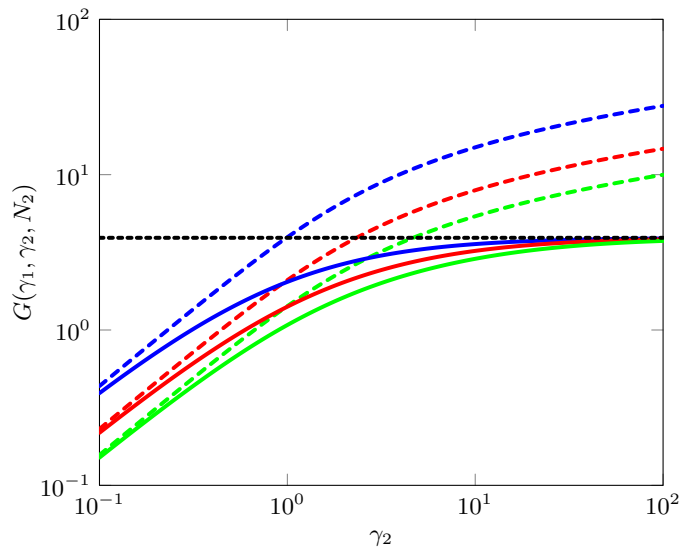
\begin{figure}
\centering
\begin{tikzpicture}
\begin{axis}[
width=0.5\textwidth,
xmin=1e-1,
xmax=1e2,
xlabel=$\gamma_2$,
xmode=log,
xminorticks=false,
ymin=1e-1,
ymax=1e2,
ylabel=${G(\gamma_1,\gamma_2,N_2)}$,
ylabel shift=-5,
ymode=log,
yminorticks=false,
]
\addplot[color=blue,dashed,no marks,line width=1.5] table
{generated_data/G_tau=50/G2_g1=0.1_N=9.dat};
\addplot[color=red,dashed,no marks,line width=1.5] table
{generated_data/G_tau=50/G2_g1=0.1_N=17.dat};
\addplot[color=green,dashed,no marks,line width=1.5] table
{generated_data/G_tau=50/G2_g1=0.1_N=25.dat};

\addplot[color=blue,no marks,line width=1.5] table
{generated_data/G_tau=50/G12_g1=0.1_N=9.dat};
\addplot[color=red,no marks,line width=1.5] table
{generated_data/G_tau=50/G12_g1=0.1_N=17.dat};
\addplot[color=green,no marks,line width=1.5] table
{generated_data/G_tau=50/G12_g1=0.1_N=25.dat};

\addplot[color=black,dotted,no marks,line width=1.5] table
{generated_data/G_tau=50/G1_g1=0.1.dat};
\end{axis}
\end{tikzpicture}
\caption{Plot of the function $G$ as defined in Eq.~\protect\eqref{envelope} as
a function of $\gamma_2$ for $\nu=0.02m$, $\gamma_1=0.1$ and $N_2=9$ (blue),
$N_2=17$ (red), $N_2=25$ (green).
The horizontal black dotted curve is for the field 1 alone, therefore
independent of $\gamma_2$.
The dashed curves are for field 2 alone, and the solid curves for the
superposition of both fields.
Note that the residual pair density is proportional to
$\exp\{-4\frac m\nu G\}$.}
\label{G two fields}
\end{figure}
Superimposing to field 1 a second field 2, i.e.\ $N=2$ in~\eqref{h}, where the
Keldysh parameter $\gamma_2$ refers to the multi-photon regime, can cause a
significant enhancement of the pair production, as pioneered in~\cite
{schutzhold_dynamically_2008,dunne_catalysis_2009} and further exemplified
in~\cite{otto_lifting_2015,otto_dynamical_2015}.
Figure~\ref{gauss super-gauss otto ass} exhibits examples for $E_1=0.2E_c$,
$\nu=0.02m$ (as in section~2) and $E_2=0.05E_c$, $N_2=25$ for pulse shape
parameters as in Fig.~\ref{gauss super-gauss otto}.
Also in the present case, the estimates~\eqref{W} and~\eqref{W'} are quite
useful, in particular for the super-Gauss (middle panel) and flat-top pulse
(right panel), while for the Gauss pulse (left panel) we meet again some
underestimate.
Quite indicative for the enhancement is the transversally integrated density,
$a_{e^-e^+}=2\pi\int\!\diff p_\perp p_\perp f(p_\perp,p_\parallel=0)$
depicted at the top of the panels, which is in all displayed cases two orders of
magnitude larger than the field 1 alone, as evident by comparison with the
analog quantities in Fig.~\ref{gauss super-gauss otto}.

In the spirit of Fig.~\ref{landscape}, one can inspect, for a rough orientation,
the pulse shape independent function $\mathcal E$ or its heart, the function
$G$, see Fig.~\ref{G two fields}.
The dotted horizontal line is for field 1 alone, while the dashed curves depict
field 2 alone, i.e.\ for $E_1=0$.
Solid curves are for the fields 1+2.
In fact, the solid curves, at fixed values of $E_{1,2}$ and $N_2$ are below the
dotted and dashed ones pointing to lowering $G$ and thus to ``lowering the
exponential suppression''.
As already observed in~\cite{orthaber_momentum_2011}, there is a certain window
where that enhancement is significant.

\section{Doubly assisted Schwinger pair production}
\begin{figure}
\centering
\begin{tikzpicture}
\begin{groupplot}[
group style={
	group size=3 by 3,
	y descriptions at=edge left,
	horizontal sep=0pt,
	vertical sep=0pt,
},
width=0.39\textwidth,
xlabel shift=-3,
ylabel shift=-5,
xmin=0,
xmax=1.5,
xtick={0, 0.3, 0.6, 0.9, 1.2, 1.5},
x tick label style={/pgf/number format/.cd, fixed, fixed zerofill, precision=1,
/tikz/.cd},
ymin=1e-12,
ymax=1e0,
ytick={1e-12, 1e-10, 1e-8, 1e-6, 1e-4, 1e-2},
xlabel=$p_\perp/m$,
ylabel={$f(t_\infty,p_\perp,p_\parallel=0)$},
ymode=log,
yminorticks=false,
]

\nextgroupplot[xlabel={}, xticklabels={},
ytick={1e-12, 1e-10, 1e-8, 1e-6, 1e-4, 1e-2, 1e0}]
\addplot[color=black,line width=1] table
{data/cut_three_fields/E1=0.1_nu=0.07_E2=0.0_N2=7_E3=0.0_N3=14.dat};
\node at (axis cs:1,1e-2) {$1$};

\nextgroupplot[xlabel={}, xticklabels={}]
\addplot[color=black,line width=1] table
{data/cut_three_fields/E1=0.0_nu=0.07_E2=0.05_N2=7_E3=0.0_N3=14.dat};
\node at (axis cs:1,1e-2) {$2$};

\nextgroupplot[xlabel={}, xticklabels={}]
\addplot[color=black,line width=1] table
{data/cut_three_fields/E1=0.0_nu=0.07_E2=0.0_N2=7_E3=0.01_N3=14.dat};
\node at (axis cs:1,1e-2) {$3$};

\nextgroupplot[xlabel={}, xticklabels={}]
\addplot[color=black,line width=1] table
{data/cut_three_fields/E1=0.1_nu=0.07_E2=0.05_N2=7_E3=0.0_N3=14.dat};
\node at (axis cs:1,1e-2) {$1+2$};

\nextgroupplot[xtick={0.3, 0.6, 0.9, 1.2}]
\addplot[color=black,line width=1] table
{data/cut_three_fields/E1=0.1_nu=0.07_E2=0.0_N2=7_E3=0.01_N3=14.dat};
\node at (axis cs:1,1e-2) {$1+3$};

\nextgroupplot
\addplot[color=black,line width=1] table
{data/cut_three_fields/E1=0.0_nu=0.07_E2=0.05_N2=7_E3=0.01_N3=14.dat};
\node at (axis cs:1,1e-2) {$2+3$};

\nextgroupplot
\addplot[color=black,line width=1] table
{data/cut_three_fields/E1=0.1_nu=0.07_E2=0.05_N2=7_E3=0.01_N3=14.dat};
\node at (axis cs:1,1e-2) {$1+2+3$};
\end{groupplot}
\tabulinesep=2pt
\node at (10, -7.5) {
\sisetup{retain-unity-mantissa=true}
\begin{tabu}{|c|c|l|}
\hline
$N$                  & field   & $a_{e^-e^+}/\lambda_C^2$\\\hline
\multirow{3}{*}{$1$} & $1$     & \num{3e-11}\\
                     & $2$     & \num{8e-8}\\
                     & $3$     & \num{1e-8}\\\hline
\multirow{3}{*}{$2$} & $1+2$   & \num{2e-5}\\
                     & $1+3$   & \num{1e-4}\\
                     & $2+3$   & \num{2e-5}\\\hline
3                    & $1+2+3$ & \num{9e-4}\\\hline
\end{tabu}};
\end{tikzpicture}
\caption{Residual transverse momentum spectra at $p_\parallel=0$ for three
different fields and their combinations as indicated.
The shape function used is the flat-top pulse.
Parameters: $E_1=0.1E_c$, $\nu=0.07m$, $E_2=0.05E_c$, $N_2=7$, $E_3=0.01E_c$,
$N_3=14$, $\nu\tramp=5\cdot2\pi$, $\nu\tflat=50\cdot2\pi$.
The table lists the transversally integrated spectra normalized to the square of
the Compton wavelength.}
\label{doubly assisted cut}
\end{figure}
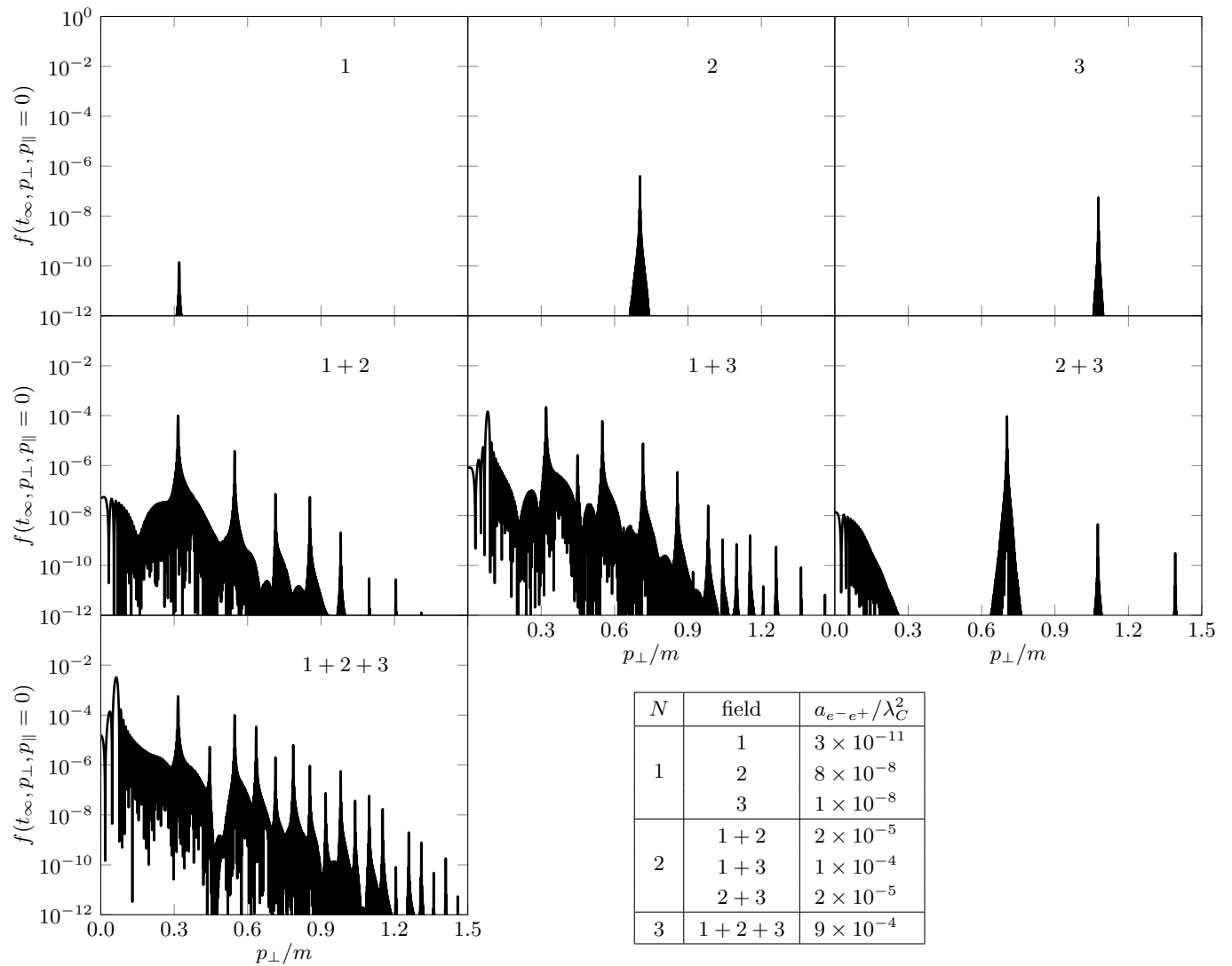
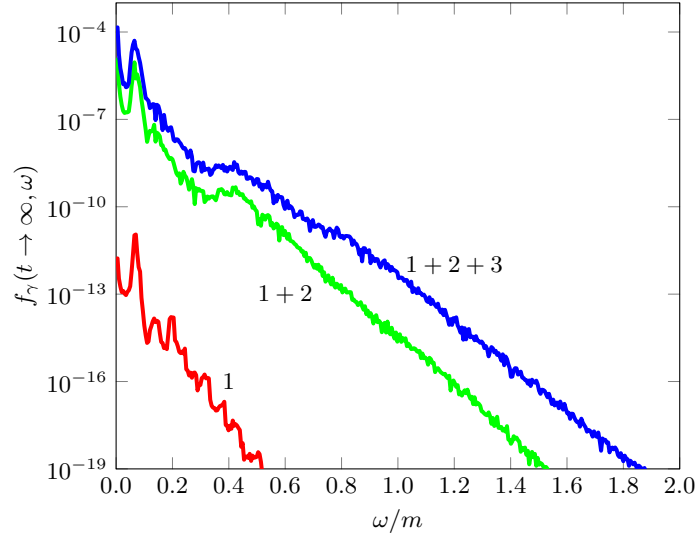
\begin{figure}
\centering
\begin{tikzpicture}
\begin{axis}[
width=0.5\textwidth,
xmin=0,
xmax=2,
xlabel=$\omega/m$,
x tick label style={/pgf/number format/.cd, fixed, fixed zerofill, precision=1,
/tikz/.cd},
ymin=1e-19,
ymax=1e-3,
ylabel=${f_\gamma(t\to\infty,\omega)}$,
ylabel shift=-3,
ymode=log,
yminorticks=false,
]
\addplot[color=red,no marks,line width=1.5] table
{data/photons/E1=0.1_nu=0.07_E2=0.0_N2=7_E3=0.0_N3=14.dat};
\node at (axis cs:1.2,1e-12) {$1+2+3$};
\addplot[color=green,no marks,line width=1.5] table
{data/photons/E1=0.1_nu=0.07_E2=0.05_N2=7_E3=0.0_N3=14.dat};
\node at (axis cs:0.6,1e-13) {$1+2$};
\addplot[color=blue,no marks,line width=1.5] table
{data/photons/E1=0.1_nu=0.07_E2=0.05_N2=7_E3=0.01_N3=14.dat};
\node at (axis cs:0.4,1e-16) {$1$};
\end{axis}
\end{tikzpicture}
\caption{Final state spectra of photons emitted with frequency $\omega$
perpendicularly to the electric field~\eqref{A} by the creation and subsequent
motion of the $e^+$$e^-$ pairs.
Field parameters and field nomenclature are as in the left column of
Fig.~\protect\eqref{doubly assisted cut}.}
\label{photons}
\end{figure}
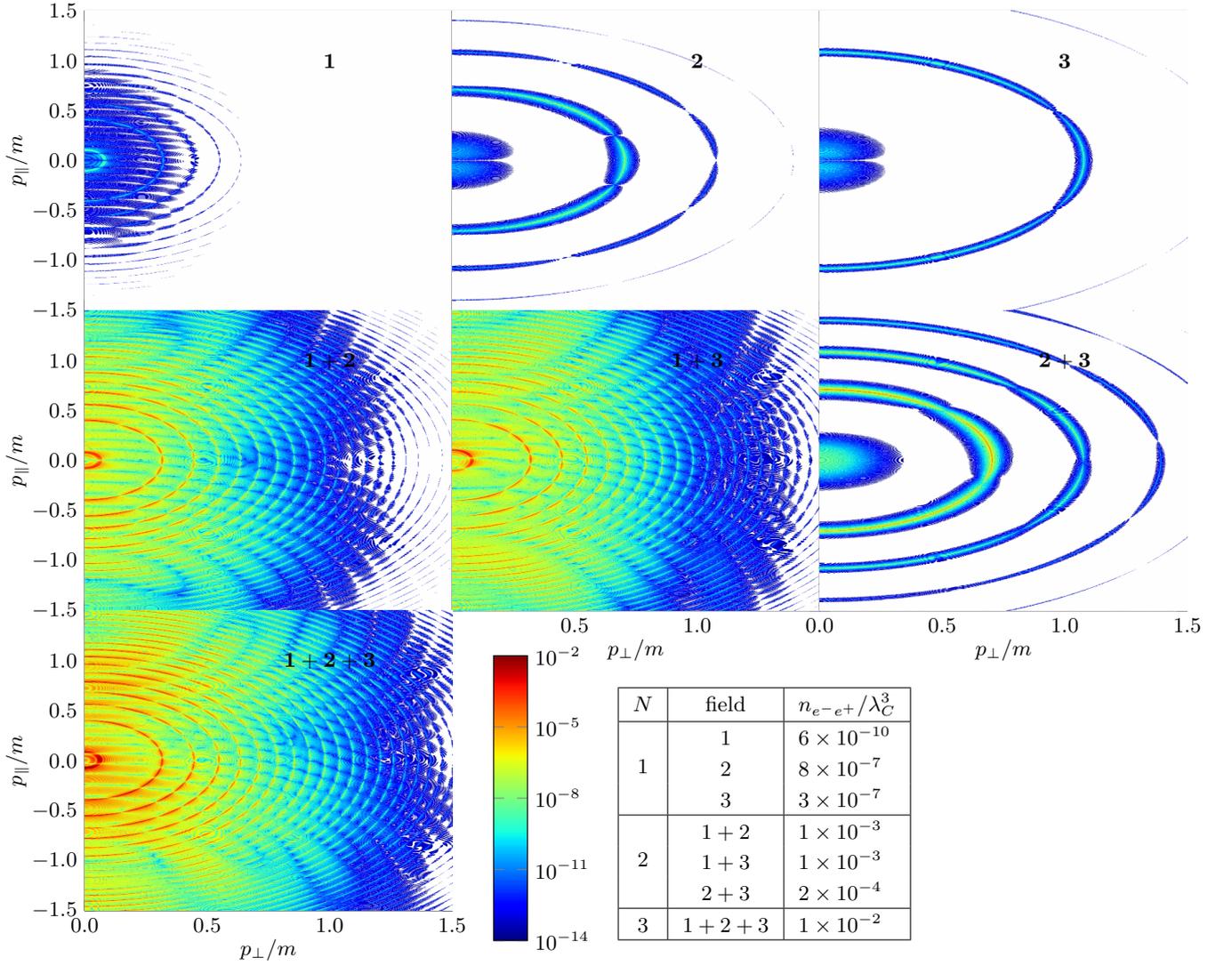
\begin{figure}
\centering
\begin{tikzpicture}
\begin{groupplot}[
group style={
	group size=3 by 3,
	y descriptions at=edge left,
	horizontal sep=0pt,
	vertical sep=0pt,
},
width=0.39\textwidth,
xlabel shift=-3,
ylabel shift=-5,
xmin=0,
xmax=1.5,
xtick={0.0, 0.5, 1.0, 1.5},
ymin=-1.5,
ymax=1.5,
ytick={-1.5, -1.0, -0.5, 0.0, 0.5, 1.0, 1.5},
xlabel=$p_\perp/m$,
ylabel={$p_\parallel/m$},
tick label style={/pgf/number format/.cd, fixed, fixed zerofill, precision=1,
/tikz/.cd},
]

\nextgroupplot[xlabel={}, xticklabels={}]
\addplot graphics[xmin=0,xmax=1.5,ymin=-1.5,ymax=1.5]
{generated_images/contours/{E1=0.1_nu=0.07_E2=0.0_N2=7_E3=0.0_N3=14}.png};
\node at (axis cs:1,1) {$\mathbf{1}$};

\nextgroupplot[xlabel={}, xticklabels={}]
\addplot graphics[xmin=0,xmax=1.5,ymin=-1.5,ymax=1.5]
{generated_images/contours/{E1=0.0_nu=0.07_E2=0.05_N2=7_E3=0.0_N3=14}.png};
\node at (axis cs:1,1) {$\mathbf{2}$};

\nextgroupplot[xlabel={}, xticklabels={}]
\addplot graphics[xmin=0,xmax=1.5,ymin=-1.5,ymax=1.5]
{generated_images/contours/{E1=0.0_nu=0.07_E2=0.0_N2=7_E3=0.01_N3=14}.png};
\node at (axis cs:1,1) {$\mathbf{3}$};

\nextgroupplot[xlabel={}, xticklabels={},
ytick={-1.5, -1.0, -0.5, 0.0, 0.5, 1.0}]
\addplot graphics[xmin=0,xmax=1.5,ymin=-1.5,ymax=1.5]
{generated_images/contours/{E1=0.1_nu=0.07_E2=0.05_N2=7_E3=0.0_N3=14}.png};
\node at (axis cs:1,1) {$\mathbf{1+2}$};

\nextgroupplot[xtick={0.5, 1.0}]
\addplot graphics[xmin=0,xmax=1.5,ymin=-1.5,ymax=1.5]
{generated_images/contours/{E1=0.1_nu=0.07_E2=0.0_N2=7_E3=0.01_N3=14}.png};
\node at (axis cs:1,1) {$\mathbf{1+3}$};

\nextgroupplot
\addplot graphics[xmin=0,xmax=1.5,ymin=-1.5,ymax=1.5]
{generated_images/contours/{E1=0.0_nu=0.07_E2=0.05_N2=7_E3=0.01_N3=14}.png};
\node at (axis cs:1,1) {$\mathbf{2+3}$};

\nextgroupplot[ytick={-1.5, -1.0, -0.5, 0.0, 0.5, 1.0}, point meta min=1e-14,
point meta max=1e-2, colormap name=myjet, colorbar, colorbar style={ymode=log,
yminorticks=false, ytick={1e-14, 1e-11, 1e-8, 1e-5, 1e-2},
height=0.95*\pgfkeysvalueof{/pgfplots/parent axis height},
anchor=south west, at={(parent axis.right of south east)},
yshift=-0.1*\pgfkeysvalueof{/pgfplots/parent axis height},
tick label style={/pgf/number format/fixed zerofill=false}}]
\addplot graphics[xmin=0,xmax=1.5,ymin=-1.5,ymax=1.5]
{generated_images/contours/{E1=0.1_nu=0.07_E2=0.05_N2=7_E3=0.01_N3=14}.png};
\node at (axis cs:1,1) {$\mathbf{1+2+3}$};
\end{groupplot}
\tabulinesep=2pt
\node at (10, -7.5) {
\sisetup{retain-unity-mantissa=true}
\begin{tabu}{|c|c|l|}
\hline
$N$                  & field & $n_{e^-e^+}/\lambda_C^3$\\\hline
\multirow{3}{*}{$1$} & $1$     & \num{6e-10}\\
                     & $2$     & \num{8e-7}\\
                     & $3$     & \num{3e-7}\\\hline
\multirow{3}{*}{$2$} & $1+2$   & \num{1e-3}\\
                     & $1+3$   & \num{1e-3}\\
                     & $2+3$   & \num{2e-4}\\\hline
3                    & $1+2+3$ & \num{1e-2}\\\hline
\end{tabu}};
\end{tikzpicture}
\caption{Color contour plots for the residual phase space distributions as in
Fig.~\ref{doubly assisted cut}.
The table lists the scaled densities.}
\label{doubly assisted contour}
\end{figure}
It has been suggested in~\cite{ilderton_nonperturbative_2015} that adding
shorter and shorter time-like inhomogeneities to the spatially homogeneous field
causes further enhancement of the pair production.
A realization of such a scenario is provided in~\cite{torgrimsson_doubly_2016}
as doubly assisted pair production.
We take the flat-top pulse shape and select $N=3$ with $E_1>E_2>E_3$ and
$N_1=1<N_2<N_3$ such that $\gamma_1$ is in the tunneling regime and
$\gamma_{2,3}$ in the multi-photon regime.
Figure~\ref{doubly assisted cut} exhibits the residual transverse momentum
spectra for the three fields alone (top row), the pairwise combination of two
fields (middle row) and the combination of all three fields (bottom left).
The table (bottom right) lists the transversally integrated spectra.
For the given parameters (see figure caption), the double assistance clearly
causes an enhancement over the single assistance case.
A similar enhancement is found in~\cite{otto_afterglow_2017} for the photons
accompanying the $e^+e^-$ pair creation as a secondary probe.
Figure~\ref{photons} exhibits an example for the above field~1 alone (curve
labelled by ``1''), for the single assistance by field~2 (curve ``1+2''), and
the double assistance (curve ``1+2+3'') as well.
The fields ``1'', ``2'' and ``3'' are the same as in Fig.~\ref{doubly assisted
cut}.
The calculations follow the first-order treatment of the quantized radiation
field coupled to the $e^+$$e^-$ pairs as presented in~\cite
{otto_afterglow_2017}.
One observes a huge assistance of the spectral photon yield $f_\gamma$ by the
single field, while the double assistance causes only a mild further
enhancement.

Figure~\ref{doubly assisted contour} returns to $e^+$$e^-$ production and
exhibits the momentum spectra as color contour plots and lists the normalized
pair densities obtained by full momentum space integration.
The latter values demonstrate the huge enhancement by multi-scale field
configurations with a clear momentum signature.

\section{Summary}
Many previous studies show that the pulse shape of spatially homogeneous, \emph
{oscillating} electric fields has a decisive impact on the residual momentum
distribution of the pairs emerging from the vacuum decay.
We provide here some approximate formula which characterizes relevant features
of the spectrum.
In particular, such pulse shapes as super-Gauss or shape functions with a clear
flat-top interval are proven to be uncovered by our formula.
We supply, besides that generic consideration, some case studies for assisted
and doubly assisted Schwinger pair creation.
These can be considered as particular realisations of multi-scale pulses.
Quite significant enhancement effects are found, also for the accompanying
photons.
While encouraging, we stress that spatial inhomogeneities counter act the
enhancement.
One has to go beyond the presently employed framework of the quantum kinetic
equations to account properly for realistic field configurations.

\bigskip\noindent\textbf{Acknowledgements:}
The authors gratefully acknowledge inspiring discussions with H.~Gies,
R.~Schützhold, R.~Alkofer, D.~B.~Blaschke and C.~Greiner.
Many thanks go to S.~Smolyansky and A.~Panferov for common work on the plain
Schwinger process.
The fruitful collaboration with R.~Sauerbrey and T.~E.~Cowan within the HIBEF
project promoted the present investigation.

The interest of one of the authors (BK) in the present topic was initiated in a
seminal physics colloquium at the TU~Dresden in 1986 by Walter Greiner, where he
surveyed the status and necessary future investigations towards understanding
the quantum vacuum.
We therefore dedicate our work to his legacy and acknowledge the collaborative
work with colleagues and friends of Walter Greiner.
\printbibliography
\end{document}

%% file: pgfsetup.tex
\usepackage{pgfplots}
\pgfplotsset{compat=1.14}
\usetikzlibrary{external}
\pgfplotsset{colormap={myjet}{[1pt]
rgb(0pt)=(0.0, 0.0, 0.5);
rgb(1pt)=(0.0, 0.0, 0.535650623885918);
rgb(2pt)=(0.0, 0.0, 0.589126559714795);
rgb(3pt)=(0.0, 0.0, 0.62477718360071299);
rgb(4pt)=(0.0, 0.0, 0.67825311942958999);
rgb(5pt)=(0.0, 0.0, 0.71390374331550799);
rgb(6pt)=(0.0, 0.0, 0.76737967914438499);
rgb(7pt)=(0.0, 0.0, 0.80303030303030298);
rgb(8pt)=(0.0, 0.0, 0.85650623885917998);
rgb(9pt)=(0.0, 0.0, 0.90998217468805698);
rgb(10pt)=(0.0, 0.0, 0.94563279857397498);
rgb(11pt)=(0.0, 0.0, 0.99910873440285197);
rgb(12pt)=(0.0, 0.0, 1.0);
rgb(13pt)=(0.0, 0.0176470588235293, 1.0);
rgb(14pt)=(0.0, 0.049019607843137254, 1.0);
rgb(15pt)=(0.0, 0.096078431372549025, 1.0);
rgb(16pt)=(0.0, 0.12745098039215685, 1.0);
rgb(17pt)=(0.0, 0.17450980392156862, 1.0);
rgb(18pt)=(0.0, 0.22156862745098038, 1.0);
rgb(19pt)=(0.0, 0.25294117647058822, 1.0);
rgb(20pt)=(0.0, 0.29999999999999999, 1.0);
rgb(21pt)=(0.0, 0.33137254901960772, 1.0);
rgb(22pt)=(0.0, 0.3784313725490196, 1.0);
rgb(23pt)=(0.0, 0.40980392156862744, 1.0);
rgb(24pt)=(0.0, 0.45686274509803909, 1.0);
rgb(25pt)=(0.0, 0.50392156862745097, 1.0);
rgb(26pt)=(0.0, 0.53529411764705859, 1.0);
rgb(27pt)=(0.0, 0.58235294117647063, 1.0);
rgb(28pt)=(0.0, 0.61372549019607847, 1.0);
rgb(29pt)=(0.0, 0.66078431372548996, 1.0);
rgb(30pt)=(0.0, 0.69215686274509802, 1.0);
rgb(31pt)=(0.0, 0.73921568627450984, 1.0);
rgb(32pt)=(0.0, 0.77058823529411768, 1.0);
rgb(33pt)=(0.0, 0.81764705882352939, 1.0);
rgb(34pt)=(0.0, 0.86470588235294121, 0.99620493358633777);
rgb(35pt)=(0.0, 0.89607843137254906, 0.97090449082858954);
rgb(36pt)=(0.034788108791903853, 0.94313725490196076, 0.93295382669196714);
rgb(37pt)=(0.060088551549652112, 0.97450980392156861, 0.9076533839342189);
rgb(38pt)=(0.098039215686274495, 1.0, 0.8697027197975965);
rgb(39pt)=(0.12333965844402275, 1.0, 0.84440227703984827);
rgb(40pt)=(0.16129032258064513, 1.0, 0.80645161290322587);
rgb(41pt)=(0.18659076533839339, 1.0, 0.78115117014547764);
rgb(42pt)=(0.22454142947501579, 1.0, 0.74320050600885512);
rgb(43pt)=(0.26249209361163817, 1.0, 0.70524984187223283);
rgb(44pt)=(0.2877925363693864, 1.0, 0.67994939911448449);
rgb(45pt)=(0.3257432005060088, 1.0, 0.6419987349778622);
rgb(46pt)=(0.35104364326375709, 1.0, 0.61669829222011385);
rgb(47pt)=(0.38899430740037944, 1.0, 0.57874762808349156);
rgb(48pt)=(0.4142947501581275, 1.0, 0.55344718532574344);
rgb(49pt)=(0.45224541429475007, 1.0, 0.51549652118912082);
rgb(50pt)=(0.49019607843137247, 1.0, 0.47754585705249841);
rgb(51pt)=(0.5154965211891207, 1.0, 0.45224541429475018);
rgb(52pt)=(0.55344718532574311, 1.0, 0.41429475015812778);
rgb(53pt)=(0.57874762808349134, 1.0, 0.38899430740037955);
rgb(54pt)=(0.61669829222011374, 1.0, 0.35104364326375714);
rgb(55pt)=(0.64199873497786197, 1.0, 0.32574320050600891);
rgb(56pt)=(0.67994939911448438, 1.0, 0.28779253636938651);
rgb(57pt)=(0.70524984187223261, 1.0, 0.26249209361163817);
rgb(58pt)=(0.74320050600885468, 1.0, 0.22454142947501621);
rgb(59pt)=(0.78115117014547741, 1.0, 0.18659076533839347);
rgb(60pt)=(0.80645161290322565, 1.0, 0.16129032258064513);
rgb(61pt)=(0.84440227703984805, 1.0, 0.12333965844402273);
rgb(62pt)=(0.86970271979759628, 1.0, 0.098039215686274495);
rgb(63pt)=(0.90765338393421868, 1.0, 0.060088551549652092);
rgb(64pt)=(0.93295382669196703, 1.0, 0.03478810879190386);
rgb(65pt)=(0.97090449082858932, 0.95933188090050858, 0.0);
rgb(66pt)=(0.99620493358633766, 0.93028322440087174, 0.0);
rgb(67pt)=(1.0, 0.88671023965141638, 0.0);
rgb(68pt)=(1.0, 0.84313725490196101, 0.0);
rgb(69pt)=(1.0, 0.81408859840232406, 0.0);
rgb(70pt)=(1.0, 0.7705156136528688, 0.0);
rgb(71pt)=(1.0, 0.74146695715323196, 0.0);
rgb(72pt)=(1.0, 0.69789397240377649, 0.0);
rgb(73pt)=(1.0, 0.66884531590413965, 0.0);
rgb(74pt)=(1.0, 0.62527233115468439, 0.0);
rgb(75pt)=(1.0, 0.58169934640522891, 0.0);
rgb(76pt)=(1.0, 0.55265068990559207, 0.0);
rgb(77pt)=(1.0, 0.50907770515613682, 0.0);
rgb(78pt)=(1.0, 0.48002904865649987, 0.0);
rgb(79pt)=(1.0, 0.4364560639070445, 0.0);
rgb(80pt)=(1.0, 0.40740740740740755, 0.0);
rgb(81pt)=(1.0, 0.36383442265795229, 0.0);
rgb(82pt)=(1.0, 0.33478576615831535, 0.0);
rgb(83pt)=(1.0, 0.29121278140886042, 0.0);
rgb(84pt)=(1.0, 0.24763979665940472, 0.0);
rgb(85pt)=(1.0, 0.21859114015976777, 0.0);
rgb(86pt)=(1.0, 0.1750181554103124, 0.0);
rgb(87pt)=(1.0, 0.14596949891067557, 0.0);
rgb(88pt)=(1.0, 0.1023965141612202, 0.0);
rgb(89pt)=(0.99910873440285231, 0.07334785766158336, 0.0);
rgb(90pt)=(0.94563279857397531, 0.029774872912127992, 0.0);
rgb(91pt)=(0.90998217468805731, 0.00072621641249104307, 0.0);
rgb(92pt)=(0.8565062388591802, 0.0, 0.0);
rgb(93pt)=(0.80303030303030321, 0.0, 0.0);
rgb(94pt)=(0.76737967914438521, 0.0, 0.0);
rgb(95pt)=(0.71390374331550821, 0.0, 0.0);
rgb(96pt)=(0.6782531194295901, 0.0, 0.0);
rgb(97pt)=(0.6247771836007131, 0.0, 0.0);
rgb(98pt)=(0.589126559714795, 0.0, 0.0);
rgb(99pt)=(0.535650623885918, 0.0, 0.0);
rgb(100pt)=(0.5, 0.0, 0.0);
}}
\pgfplotsset{colormap={myjet_r}{[1pt]
rgb(0pt)=(0.5, 0.0, 0.0);
rgb(1pt)=(0.535650623885918, 0.0, 0.0);
rgb(2pt)=(0.589126559714795, 0.0, 0.0);
rgb(3pt)=(0.62477718360071299, 0.0, 0.0);
rgb(4pt)=(0.67825311942958999, 0.0, 0.0);
rgb(5pt)=(0.71390374331550799, 0.0, 0.0);
rgb(6pt)=(0.76737967914438499, 0.0, 0.0);
rgb(7pt)=(0.80303030303030298, 0.0, 0.0);
rgb(8pt)=(0.85650623885918009, 0.0, 0.0);
rgb(9pt)=(0.90998217468805709, 0.0007262164124910357, 0.0);
rgb(10pt)=(0.94563279857397509, 0.029774872912127923, 0.0);
rgb(11pt)=(0.99910873440285208, 0.073347857661583263, 0.0);
rgb(12pt)=(1.0, 0.10239651416122014, 0.0);
rgb(13pt)=(1.0, 0.14596949891067537, 0.0);
rgb(14pt)=(1.0, 0.17501815541031238, 0.0);
rgb(15pt)=(1.0, 0.21859114015976769, 0.0);
rgb(16pt)=(1.0, 0.24763979665940458, 0.0);
rgb(17pt)=(1.0, 0.29121278140885992, 0.0);
rgb(18pt)=(1.0, 0.33478576615831523, 0.0);
rgb(19pt)=(1.0, 0.36383442265795213, 0.0);
rgb(20pt)=(1.0, 0.4074074074074075, 0.0);
rgb(21pt)=(1.0, 0.43645606390704428, 0.0);
rgb(22pt)=(1.0, 0.48002904865649976, 0.0);
rgb(23pt)=(1.0, 0.5090777051561366, 0.0);
rgb(24pt)=(1.0, 0.55265068990559174, 0.0);
rgb(25pt)=(1.0, 0.59622367465504733, 0.0);
rgb(26pt)=(1.0, 0.62527233115468395, 0.0);
rgb(27pt)=(1.0, 0.66884531590413954, 0.0);
rgb(28pt)=(1.0, 0.69789397240377637, 0.0);
rgb(29pt)=(1.0, 0.74146695715323152, 0.0);
rgb(30pt)=(1.0, 0.77051561365286858, 0.0);
rgb(31pt)=(1.0, 0.81408859840232395, 0.0);
rgb(32pt)=(1.0, 0.8431372549019609, 0.0);
rgb(33pt)=(1.0, 0.88671023965141615, 0.0);
rgb(34pt)=(0.99620493358633766, 0.93028322440087152, 0.0);
rgb(35pt)=(0.97090449082858932, 0.95933188090050836, 0.0);
rgb(36pt)=(0.93295382669196703, 1.0, 0.034788108791903867);
rgb(37pt)=(0.90765338393421868, 1.0, 0.060088551549652133);
rgb(38pt)=(0.86970271979759639, 1.0, 0.098039215686274536);
rgb(39pt)=(0.84440227703984805, 1.0, 0.1233396584440228);
rgb(40pt)=(0.80645161290322576, 1.0, 0.16129032258064518);
rgb(41pt)=(0.78115117014547741, 1.0, 0.18659076533839347);
rgb(42pt)=(0.74320050600885512, 1.0, 0.22454142947501587);
rgb(43pt)=(0.70524984187223261, 1.0, 0.26249209361163828);
rgb(44pt)=(0.67994939911448449, 1.0, 0.28779253636938651);
rgb(45pt)=(0.64199873497786197, 1.0, 0.32574320050600891);
rgb(46pt)=(0.61669829222011374, 1.0, 0.3510436432637572);
rgb(47pt)=(0.57874762808349134, 1.0, 0.3889943074003796);
rgb(48pt)=(0.55344718532574322, 1.0, 0.41429475015812767);
rgb(49pt)=(0.5154965211891207, 1.0, 0.45224541429475024);
rgb(50pt)=(0.4775458570524983, 1.0, 0.49019607843137264);
rgb(51pt)=(0.45224541429475007, 1.0, 0.51549652118912093);
rgb(52pt)=(0.41429475015812767, 1.0, 0.55344718532574333);
rgb(53pt)=(0.38899430740037944, 1.0, 0.57874762808349156);
rgb(54pt)=(0.35104364326375703, 1.0, 0.61669829222011396);
rgb(55pt)=(0.3257432005060088, 1.0, 0.64199873497786231);
rgb(56pt)=(0.2877925363693864, 1.0, 0.6799493991144846);
rgb(57pt)=(0.26249209361163817, 1.0, 0.70524984187223294);
rgb(58pt)=(0.2245414294750161, 1.0, 0.74320050600885501);
rgb(59pt)=(0.18659076533839336, 1.0, 0.78115117014547775);
rgb(60pt)=(0.16129032258064513, 1.0, 0.80645161290322598);
rgb(61pt)=(0.12333965844402273, 1.0, 0.84440227703984838);
rgb(62pt)=(0.098039215686274495, 1.0, 0.86970271979759661);
rgb(63pt)=(0.060088551549652092, 0.97450980392156861, 0.90765338393421902);
rgb(64pt)=(0.03478810879190386, 0.94313725490196076, 0.93295382669196736);
rgb(65pt)=(0.0, 0.89607843137254906, 0.97090449082858976);
rgb(66pt)=(0.0, 0.86470588235294121, 0.99620493358633799);
rgb(67pt)=(0.0, 0.81764705882352939, 1.0);
rgb(68pt)=(0.0, 0.77058823529411768, 1.0);
rgb(69pt)=(0.0, 0.73921568627450984, 1.0);
rgb(70pt)=(0.0, 0.69215686274509802, 1.0);
rgb(71pt)=(0.0, 0.66078431372549018, 1.0);
rgb(72pt)=(0.0, 0.61372549019607847, 1.0);
rgb(73pt)=(0.0, 0.58235294117647052, 1.0);
rgb(74pt)=(0.0, 0.53529411764705881, 1.0);
rgb(75pt)=(0.0, 0.4882352941176471, 1.0);
rgb(76pt)=(0.0, 0.45686274509803926, 1.0);
rgb(77pt)=(0.0, 0.40980392156862744, 1.0);
rgb(78pt)=(0.0, 0.3784313725490196, 1.0);
rgb(79pt)=(0.0, 0.33137254901960789, 1.0);
rgb(80pt)=(0.0, 0.30000000000000004, 1.0);
rgb(81pt)=(0.0, 0.25294117647058822, 1.0);
rgb(82pt)=(0.0, 0.22156862745098038, 1.0);
rgb(83pt)=(0.0, 0.17450980392156912, 1.0);
rgb(84pt)=(0.0, 0.12745098039215685, 1.0);
rgb(85pt)=(0.0, 0.096078431372549011, 1.0);
rgb(86pt)=(0.0, 0.049019607843137303, 1.0);
rgb(87pt)=(0.0, 0.01764705882352946, 1.0);
rgb(88pt)=(0.0, 0.0, 1.0);
rgb(89pt)=(0.0, 0.0, 0.99910873440285231);
rgb(90pt)=(0.0, 0.0, 0.94563279857397531);
rgb(91pt)=(0.0, 0.0, 0.90998217468805731);
rgb(92pt)=(0.0, 0.0, 0.8565062388591802);
rgb(93pt)=(0.0, 0.0, 0.80303030303030321);
rgb(94pt)=(0.0, 0.0, 0.76737967914438521);
rgb(95pt)=(0.0, 0.0, 0.71390374331550821);
rgb(96pt)=(0.0, 0.0, 0.6782531194295901);
rgb(97pt)=(0.0, 0.0, 0.6247771836007131);
rgb(98pt)=(0.0, 0.0, 0.589126559714795);
rgb(99pt)=(0.0, 0.0, 0.535650623885918);
rgb(100pt)=(0.0, 0.0, 0.5);
}}
\tikzset{align at top/.style={baseline=(current bounding box.north)}}
\tikzset{every picture/.append style={line join=round,line cap=round,}}
\usetikzlibrary{external}
\usetikzlibrary{decorations,decorations.pathmorphing,decorations.markings}
\usetikzlibrary{fpu,matrix,arrows,positioning,fadings,shapes,plotmarks,fit}
\usetikzlibrary{pgfplots.groupplots}
\usetikzlibrary{cd}